# A cognitive radio ad hoc network's based disaster management scheme with efficient spectrum management, collaboration and interoperability

Noman Islam
Iqra University
Karachi, Pakistan
noman.islam@gmail.com

Ghazala Shafi Sheikh
Department of Computing and Technology
Indus University
Karachi, Pakistan
ghazala.shafi@indus.edu.pk

Zeeshan Islam
ALADIN Solutions

*Abstract*— In this work, a disaster management scheme based on cognitive radio ad hoc network (CRAHN) has been presented. Disaster management has been a big problem for mankind for years. However, still not much research work has been presented on this problem. Technology has been employed in past few years to address this problem. Cognitive radio ad hoc network presents a viable solution for disaster management. It can be deployed rapidly without infrastructure and it solves the spectrum scarcity and congestion issues that arise during disaster. This paper presents a novel solution for disaster management. It provides a multi-layer perceptron (MLP) based disaster detection scheme based on WSN. To solve the spectrum scarcity problem, a MLP based spectrum management scheme has been proposed. In order to ensure collaboration among rescue workers during disaster, a novel service discovery scheme has been proposed. To ensure interoperability during communication, XML format has been recommended. A real-time GUI has been proposed to provide shared situation awareness to rescue workers and enabling better decision making. The proposed approach has been implemented in NS-2 simulator. The results show accurate disaster detection, efficient spectrum usage, and interoperability and collaboration among nodes with reduced latency.

INTRODUCTION

A disaster is the outcome of sudden event (such as natural calamity or man-made event) that hampers the normal functioning and the situation can't be resumed without external help. Various types of disasters are possible such as natural disaster, environmental disaster, complex emergencies or pandemic emergencies [23]. During the past years incidence of disasters have increased tremendously due to various factors such as environmental changes, technology failure, malfunctioning of machinery and displacement of people. Irrespective of disaster, an approach to disaster management is pivotal for curbing the loss that can occur due to disaster.

The progression in technology in recent years has enabled the use of Information and Communication Technology (ICT) for management of disaster. A number of solutions have thus been proposed as discussed in literature review section. By using the ICT technology, disasters can be detected in advance or in real-time. The warnings can thus be propagated to effected region using radio, television, cellular network and internet. Relief operations can also be supported using the technology such as using Unmanned Aerial Vehicles (UAV), wireless and mobile ad hoc network.

This research work proposes a novel solution for disaster management based on cognitive radio ad hoc network (CRAHN). A CRAHN is defined as a collection of mobile radio nodes that are organized on the fly to form a network. It is to be noted that the nodes in CRAHN can sense and use the radio frequency spectrum available in its surroundings.

**Paper's contribution:**

The proposed framework employs a four prong solution i.e. monitoring, response, assessment and recovery as shown in Figure 1. The disaster prone region is continuously monitored for any happenings using sensor network. Once disaster happened, the relevant people are disseminated the information very quickly using the ad hoc network. The assessment of the disaster intensity and impact is determined and appropriate recovery is initiated.

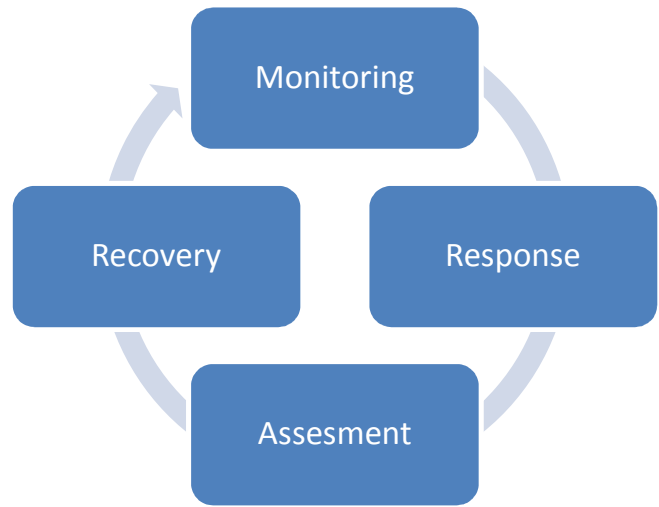

Fig. 1. A four prong solution for disaster management

To employ the proposed CRAHN based solution, several issues need to be addressed:

- **Real-time detection of disaster:** Any disaster must be detected on real-time to minimize the losses. The proposed approach deploys sensor network on site and then uses an ANN based disaster detection technique for accurate detection of disaster.

- **Informing the right personnel:** Once a disaster has been detected, it should be communicated to appropriate person. However, most of the infrastructure at disaster site has been collapsed. To communicate with outside world, a gateway node is required. In this paper, a novel gateway discovery scheme is proposed. A gateway node is selected using proposed scheme and then using the gateway node, outside world is informed about the disaster.

- **Managing/ sharing the available spectrum:** At the time of disaster, the whole communication infrastructure is wiped out. The available infrastructure is not sufficient and suffers from congestion. In order for enabling communication among rescue workers, a CRAHN is deployed at the disaster site. The licensed spectrum of other users (called primary user) is used for communication at the time of primary user's inactivity. For this purpose, spectrum holes are searched. A spectrum hole at a particular instant of time is a communication channel in which primary user is not transmitting. Since more than one spectrum hole can be available, an ANN based approach is proposed to select the best available spectrum holes based on past activity of primary user.

- **Collaboration among Rescue Teams:** It has been observed that lack of clear plan and coordination among rescue team results in reduction in effectiveness of emergency response, desperation and more casualties [3]. To address this problem, shared situation awareness is essential. Runtime services are provided for collaboration at disaster site. A real-time GUI is proposed to provide situation awareness to the rescue teams and collaboration among nodes.

- **Interoperability:** One of the most important issue during the time disaster is interoperability among nodes [4]. In order for the rescue nodes to communicate, they must agree to a uniform standard. To enable syntactic interoperability among various interacting entities, an XML based approach has been used. Software ontology has been incorporated to ensure semantic interoperability among rescue workers.

Rest of the sections of this paper discuss the proposed solution. It starts with discussion on related work and builds the hypothesis. The following sections present the work which is followed with results. The paper concludes with recommendations and future work.

LITERATURE REVIEW

In initial days, manual techniques were employed for disaster management. Among the manual techniques were dogs, cameras, sound detectors, shouting and listening to a trapped persons [3]. Such efforts are proven to be not very effective as well as time consuming. Recently, ICT technologies have been tried to manage disasters. InMeissner et al. [6] advocated for an integrated solution for disaster response and recovery. Xu and Zlatanova [4] have proposed an ontology-based architecture for ensuring interoperability at the time of emergency scenarios. The proposed architecture enables integration of information such that user's queries can be automatically answered during disaster. Fiedrich and Burghardt [7] have discussed the role of software agents during disaster. The various properties of agents such as autonomy, environment sensing and BDI model can be used for data acquisition, coordination and autonomous decision making. Therese et al. [8] have proposed a novel mobile-based approach for coordination at the time of disaster. Based on genetic algorithms, the travelling sales man problem is solved such that an optimum route is determined for a rescuer to entertain maximum number of people and cover the maximum area. SENDROM [3] uses sensor nodes for management of disaster. Two types of sensor nodes i.e. s-nodes and i-nodes are deployed during disaster. There are c-nodes that are deployed close to emergency centers. The c-nodes are queries by rescuers to find a person in disaster area.

Workpad is an integrated architecture [10] for disaster management. The proposed architecture comprises numerous front-end layers called first responders and back-end layers that work in tandem for responding to disasters. In some of the literature, various ad hoc network-based solutions have been proposed for managing disaster [2, 11, 12, 13]. DistressNet is an ad hoc network-based solution for situation management presented by George et al. [9]. The solution is based on a collaborative approach to spectrum sensing, toplogy-driven routing and localization of resources MobileMap[14] is a collaborative platform for communication among firefighters that also supports decision making. In [15], authors have proposed various architectures of sensor networks for detection of different types of disasters such as earthquake, flood and forest fire detection etc.

Having talked about various disaster management solution based on conventional approaches, the next few lines talk about disaster management operations specifically executed using CRAHN. A novel and robust CRAHN solution-based on erection for clusters for responding to disasters have been presented in [16]. The proposal is based on a 3-tier resource management scheme and specifically handles post-disaster situation.

Khayami et al. [17] advocated for a solution for whole city based on CRAHN. Thesolution proposed that smart grid communication is established for resolving communication problem in a disaster solution. Sun et al. [18] have employed a rapid solution for disaster management based on cognitive radio vehicles

Based on our study, it has been identified that the work on disaster management is very pre-mature. Specifically, there are only few solutions that employ CRAHN for management of disaster. These solutions primarily address the issues of formation of a communication infrastructure for managing disaster. But, they don't talk about aspects such as spectrum management, collaboration and interoperability among rescue workers at the time of disaster. CRAHN is a viable candidate for disaster management because it can be spontaneously deployed at the time of disaster and it enables optimum spectrum usage.

In this paper, a novel CRAHN based disaster response system has been proposed, extending the earlier work presented in [24]. It proposes solutions for disaster detection, optimal spectrum usage, collaboration and interoperability among rescue workers. Even though, aspects such as spectrum management have been considered individually in research literature, but they have not been considered in the domain of disaster management. The novelty of proposed work lies in advocating for a comprehensive disaster response approach with individual solutions targeted for the domain of disaster management. Another novelty of proposed approach is that unlike most of the spectrum management approaches, the proposed spectrum management considers the usage history of primary user for selection of appropriate spectrum hole.

PROPOSED DISASTER MANAGEMENT SYSTEM

In this section, the proposed disaster response system is discussed. The various phases of disaster management are deliberated and the proposal is discussed in detail.

*Disaster detection*

Fig. 2a illustrates the proposed solution for disaster detection running at the disaster site [24]. Various types of seismic/ smoke sensors are installed at the disaster site. The observations of the sensors served as features for a multi-layer perceptron (MLP). This information is recorded in a context database and serves as input to MLP. The MLP is pre-trained on historical records related to disaster and can be used to make future prediction related to disaster. The MLP performs a two-class classification problem based on contextual data i.e. whether a disaster has happened or not. Fig. 2b shows the sensor network deployed at the disaster site. The sensors are organized in the form of clusters. The sensor data is passed to cluster heads. The cluster heads passed sensor data to sink nodes that are ultimately passed on to disaster detection node.

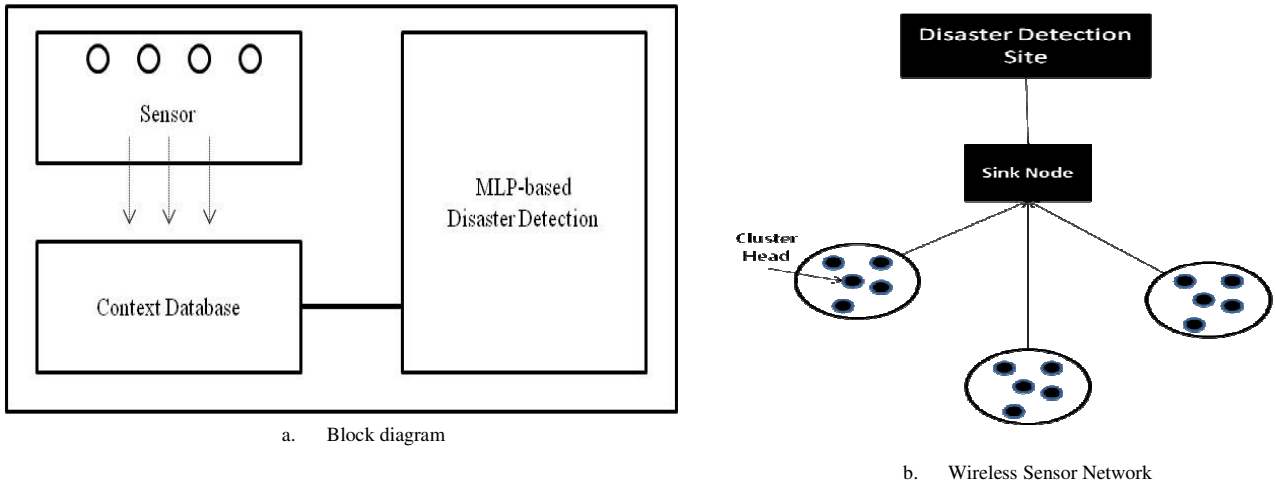

a. Block diagram

b. Wireless Sensor Network

Fig. 2. Proposed Disaster Detection System

Fig 3 shows the pseudo-code for proposed disaster detection scheme. When the system starts, it trains the ANN classifier. The system also obtains a reference to context model. It then periodically checks if disaster has happened or not. In case of a disaster, rescue work is started.

```
public class DisasterDetection {
  private ANN a = new ANN();
  public static final int DISASTER_HAPPENED = 101;
  public static final int DISASTER_NOT_HAPPENED = 102;
  private ContextModel cm;
  public DisasterDetection() {
    cm = Runtime.getContextModel();
    a.train();
    this.monitor();
  }
  pubic void monitor() {
    while(true) {

      if(a.classify(cm) == DisasterDetection. DISASTER_HAPPENED) {
        //do rescue worker
      }
      Thread.sleep(10 * 1000);
    }
  }
}
```

Fig. 3. Pseudo-code for proposed disaster detection

*Spectrum Management*

Once the disaster has been detected, the response work is required to be initiated. In a disastrous situation, most of the communication infrastructure is destroyed [2]. However, the rescue teams operating in disaster place are required to be in constant communication in order to perform effective rescue operation. An infrastructure-less network that is operating in Industrial, Scientific and Medicine (ISM) band without the help of a prior infrastructure is desirable in situation. [19].

We propose a CRAHN-based solution for managing communication in disaster scenario. Since, the CRAHN is an infrastructure-less network the network is formed as follows. The nodes operating in region will use beacon messages to detect their neighbors that can be communicated using single-hop. Using connected component analysis, all the nodes that are connected (either using single hop or multi-hops) will form an ad hoc network. Having the CRAHN established, the secondary users can operate on spectrum of primary users. This approach will solve the problem of scarcity of spectrum.

Fig. 4 illustrates the block diagram of proposed spectrum management solution. There will cognitive radios deployed at every nodes of the network and will be used to detect spectrum holes. A spectrum hole is a portion of spectrum that is not in used by primary user at a particular time. The system logs primary user's spectrum usage $l_{pi,n}$ (the duration of last $n$ sessions of every primary user) along with the other details of primary user's spectrum i.e. signal strength $s_i$ and mobility $m_i$. The spectrum log serves as a feature for an MLP-based spectrum manager. The proposed MLP uses machine learning techniques to select the best spectrum hole. Besides, there is a situation database maintained at every node. This is used for specification of the real-time update of the situation at disaster site (see section on real-time data delivery). The information contained in situation database can be visualized using Real-time GUI.

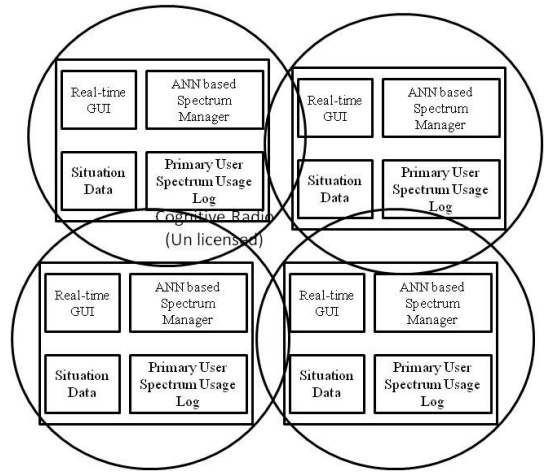

Fig. 4. Cognitive Radio based communication during disaster [24]

*Response Initiation*

The system now moves into the response phase. This work proposes an autonomous alert system that communicates the disaster information to appropriate person in the call tree. However, as already discussed, the internet infrastructure is completely destroyed. There must be some mechanism to identify a gateway node through which communication with outside world can be restored. A service discovery scheme has been proposed based on Islam and Sheikh [20]. A gateway discovery request is initiated which is propagated on the network. Ultimately, a gateway node is discovered on the network that can be invoked to handle communication with outside world and contacting the appropriate person in the call tree.

*Managing disaster*

The management of disaster comprises starting various rescue services on the network that are available to be consumed by rescue workers and affected person at the disaster site. These services can be discovered using the proposed discovery algorithm and then invoke to perform various tasks. For instance, an injured person can consume a rescue 112 service to contact rescue team and communicate its whereabouts, seeking help to rescue from the place. A rescue team member can similarly obtain a real-time status of current situation at various places or to inquire about explosive gases in the air. [6]. The services information is used by nodes to extract situation data which is then saved in situation database.

Fig 5a shows the GUI used for managing services by rescue workers. As will be shown in next section, the services are advertised by every node. The rescue workers can place their services in the network. A rescue worker can ask for services by posing a query as shown in Fig 5b

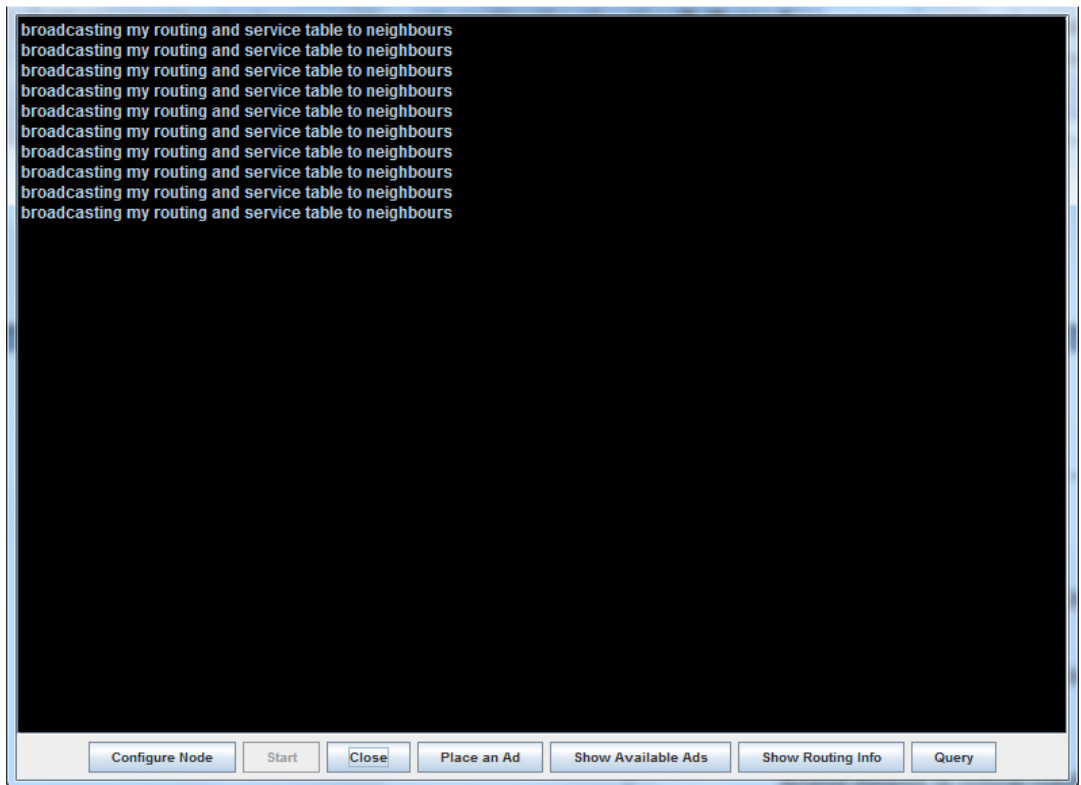

a. GUI for managing disaster

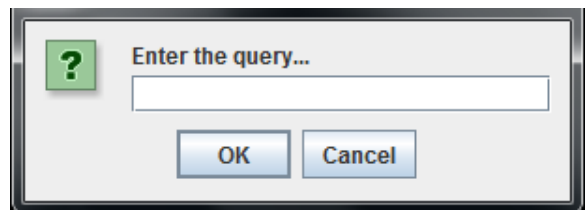

b. Service request by a rescue worker

Fig. 5. Managing disaster using proposed GUI

*Service discovery algorithm*

The proposed service discovery algorithm is based on periodic dissemination of services information by nodes along with the corresponding routes. The nodes receiving the disseminated message can save it into local cache which can then be served locally upon a request from the same node. However, if a request can't be served locally, it is then served by floating on the network using AODV routing protocol as discussed in Islam and Sheikh [20]. The intermediate nodes propagates the service request until it reaches to the target node which then responds back with the information about the desired service. A corresponding route towards the source is ultimately established thus saving an additional pass that might be required for route determination.

*Real-time data delivery*

The real-time situation awareness is an important aspect of any disaster management solution. If the information can't be obtained in real-time, it can lead to loss of lives and other damages. [9]. Unfortunately, the current solutions available in literature for data management employ manual tables and charts. The information is updated manually by exchanges of verbal messages communicated via wireless radios. Certainly, this manual approach is not viable. The manual situation can't support real-time information update nor does any real-time decision making can be performed. This work proposes a real-time GUI for rescue workers running on their computing machinery (tablet/ laptop). This GUI can be used to get a snapshot of current situation at the rescue site. Fig 6 shows the proposed real-time GUI. This GUI can be used to make real-time decision and appropriate action relevant to situation.

| Location | Situation | TimeStamp | ShortMessage |
|---|---|---|---|
| 24.8614620, 67.0099390 | Red | 20052015201820 | Injured Persons in critical condition |
| 24.8615620, 67.0039390 | Green | 20052015200820 | Rescue Work successfully done |
| 24.8614220, 67.0094390 | Yellow | 20052015200720 | Rescue operation going on |

Fig. 6. Snapshot of Real-time GUI

An important aspect for any disaster management situation is ensuring the interoperability among the nodes. There can be syntactic and semantic differences among the representation of information. However, this problem can be addressed by using a common XML format for maintain and disseminating situation dada. Fig 7 shows a sample message. It contains information about latitude, longitude, situation, timestamp, short message, long message and the ontology required to understand the semantics.

```
<?xml>
<XML>
        <Location>
                <Latitude>24.8614220<Latitude>
                <Longitude>67.0094390 <Longitude>
        </Location>
        <Situation>Red</Situation>
        <TimeStamp>20052015201820</TimeStamp>
        <ShortMessage>
            Injured Persons in critical condition
        </ShortMessage>
        <LongMessage>
             Injured Persons in critical condition stucked.
              Immediate help required. Bring cranes, cutters
              along with you
        </LongMessage>
        <Ontology>
              Safety
        </Ontology>
</XML>
```

Fig. 7. Sample XML Message

IMPLEMENTATION DETAILS & RESULTS

The proposed disaster management solution has been validated on NS-2 [21]. A cognitive radio ad hoc network is established over area of 1000 × 1000 m$^2$. For simulation of disaster, earthquake data available at [22] is used. Table 1 shows the various parameters used for simulation.

Table 1: Simulation Parameters

| Parameter | Value |
|---|---|
| Simulation time | 500 s |
| Simulation Area | 1000 m×1000 m |
| MAC layer | IEEE 802.11 |
| Routing algorithm | AODV |
| Path loss model | Free space |
| Mobility model | Random way point |

Fig. 8 analyzes the false negative alarm rate (missed disaster) and response time for disaster detection with various cluster nodes The various sensors at site is aggregated in different clusters whose data to sink and then to ANN running at disaster detection site. The average false negative alarm rate is considerably low about 1.76%. As the number of clusters at sensing site is increased the false negative alarm rate is reduced further. However, as can be seen from Fig 8b, the average response time increases with the number of cluster nodes. This is due to more amount of data to be processed by ANN based disaster detection system.

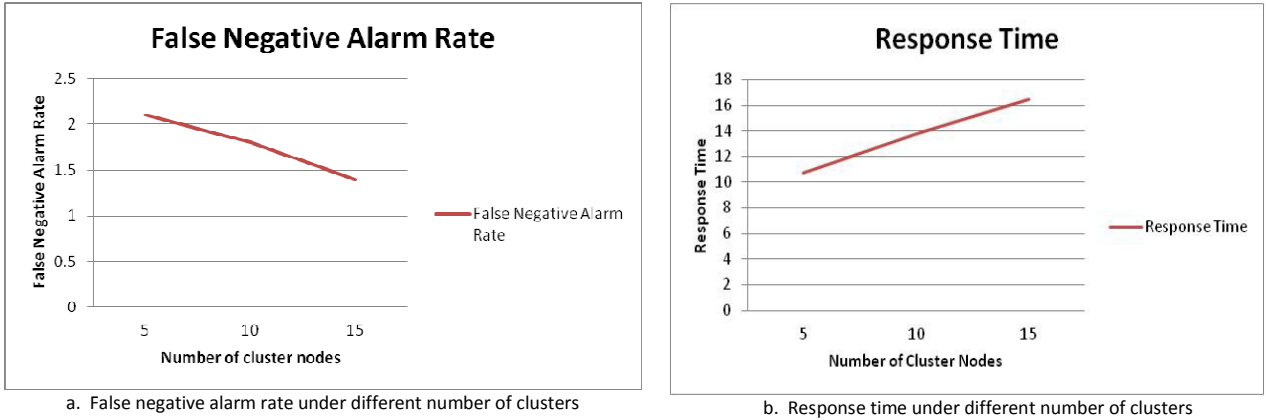

a. False negative alarm rate under different number of clusters

b. Response time under different number of clusters

Fig. 8. False Negative Alarm Rate and Response Time for proposed disaster detection scheme

Fig. 9 summarizes the spectrum switching time of secondary users (SU) for various numbers of primary users (PU). The *spectrum switching time* is defined as the amount of time during which primary user's spectrum is available to a secondary user. After this time period, the secondary user is bound to switch to spectrum of another primary user. The average spectrum switching time is 1.3 sec.

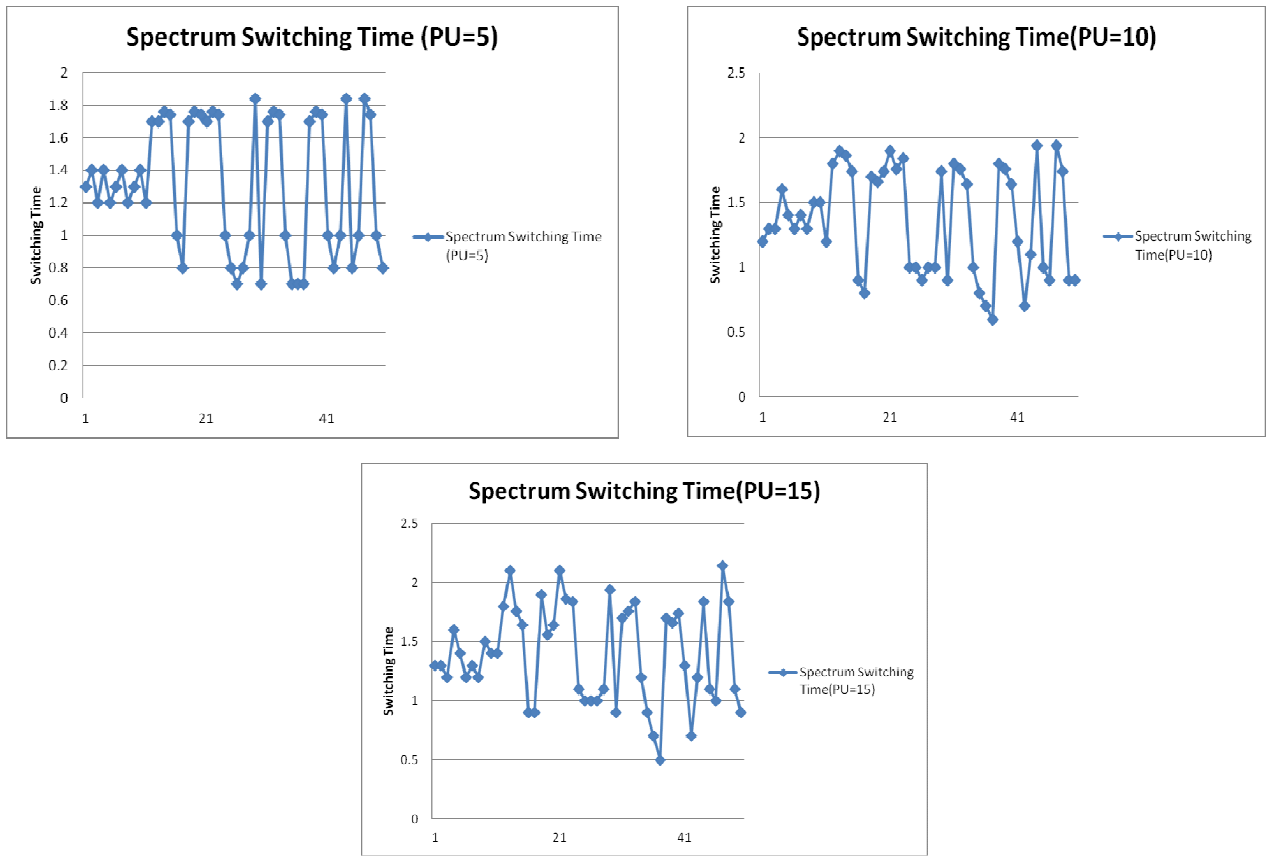

Fig. 9. Spectrum switching time for various numbers of primary users

Fig. 10 compares the spectrum switching time when PU's activity is considered by ANN based spectrum management. As discussed earlier, the ANN based spectrum manager considers PU's usage log along with signal strength and mobility of PU. The figure compares the result with the case when PU's past activity is not considered. It can be seen that by employing PU's past activity, better results in terms of spectrum usage is obtained.

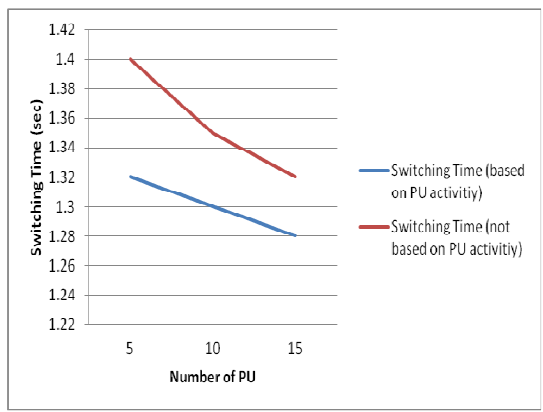

Fig. 10. Comparison of spectrum switching time based on PU's past activity Vs not based on PU's past activity

Concluding this section, Fig. 11a analyzes the latency of proposed service discovery scheme. The number of services is set to 10 while the number of nodes is set to 50. Fig 11b shows the latency for various number of computing nodes and services available on the network.

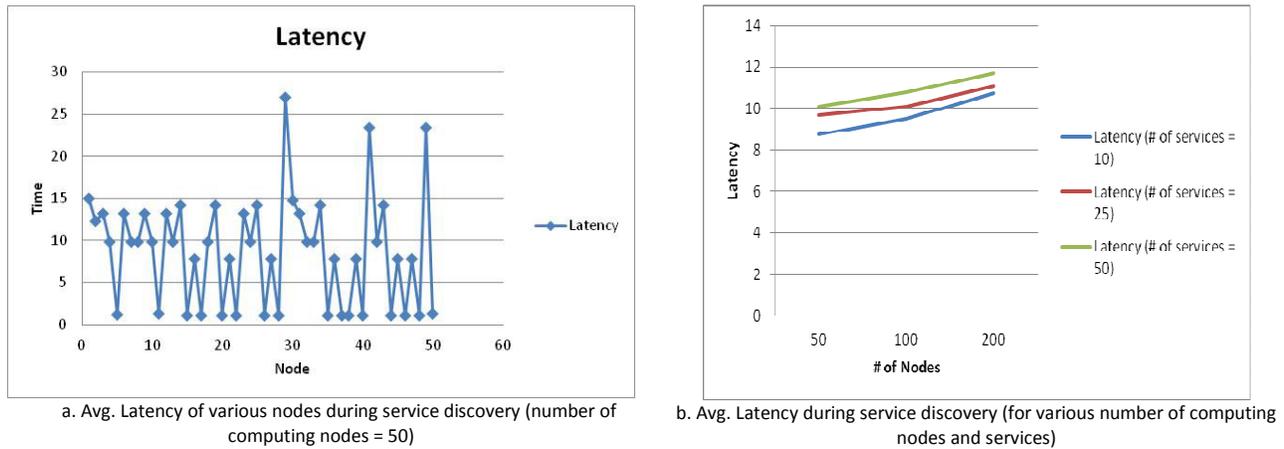

a. Avg. Latency of various nodes during service discovery (number of computing nodes = 50)

b. Avg. Latency during service discovery (for various number of computing nodes and services)

Fig. 11. Average latency of various computing nodes

ACKNOWLEDGMENTS

CONCLUSION

This work presents a disaster management solution based on cognitive radio ad hoc network (CRAHN). The proposition is based on a MLP-based disaster detection and spectrum solution along with a novel service discovery scheme. The proposed approach has been implemented in NS-2 that shows satisfactory disaster detection and spectrum management performance. The future work comprises the deployment of proposed approach on real-world environment and further analysis with large number of nodes.